\documentclass[preprint,12pt]{elsarticle}

\usepackage{amssymb}
\usepackage{url}
\usepackage{booktabs} 

\usepackage{hyperref}
\usepackage{caption}
\usepackage{rotating}
\usepackage{graphicx}
\usepackage{multirow}
\usepackage{array}
\usepackage{amsmath}

\begin{document}

\begin{frontmatter}



\author[1]{Muhammad Raees}
\ead{raees.se@must.edu.pk}
\affiliation[1]{organization={Mirpur University of Science and Technology},
            city={Mirpur},
            state={AJK},
            country={Pakistan}}
\author[1]{Afzal Ahmed}
\ead{afzal.se@must.edu.pk}

\title{Context-Aware Agent-based Model for Smart Long Distance Transport System}

\begin{abstract}
Long-distance transport plays a vital role in the economic growth of countries. However, there is a lack of systems being developed for monitoring and support of long-route vehicles (LRV). Sustainable and context-aware transport systems with modern technologies are needed. We model for long-distance vehicle transportation monitoring and support systems in a multi-agent environment. Our model incorporates the distance vehicle transport mechanism through agent-based modeling (ABM). This model constitutes the design protocol of ABM called Overview,
Design and Details (ODD). This model constitutes that every category of agents is offering information as a service. Hence, a federation of services through protocol for the communication between sensors and software components is desired. Such integration of services supports monitoring and tracking of vehicles on the route. The model simulations provide useful results for the integration of services based on smart objects.
\end{abstract}


\end{frontmatter}


\section{Introduction}
The smart transport sector positively impacts the economy which is vital for growth, social welfare, and development \cite{yang2010survey}. Thus, detecting changes and forecasting future freight flows and freight transport demand is a task of great importance \cite{vermesan2022internet}.
Long-distance freight transfer booms the economy and generates a wide range of employment for common people and business opportunities for investors \cite{yang2010survey}.
However, in third-world countries like Pakistan where there are not many systems that are developed for the tracking or monitoring of long distance vehicles, this transportation medium is not very productive \cite{raees2021context}. 
Thus, harming the economy due to long delays in the transportation of goods. Some of the common problems faced by long-distance vehicles range from small breakdowns to stolen goods or vehicles. 
Other effects like weather, lack of knowledge about the rest areas, fuel stations, vehicle service areas, road police or ambulance services for any emergency are lacking due to minimum or no communication with the vehicle owner company and other help services. The problems also extend to local public authorities causing blockage on roads on the local, regional, or national level to mega economic issue of not reaching goods from one part to another.

Therefore, a need arises for a sustainable transport system incorporated with modern technologies that is capable of fulfilling
the transport needs of society. At the same time, there is a need to hinder the negative effects of long-distance transportation.
By incorporating various types of technologies currently available and other infrastructural measures, it is often possible to influence how these actions are selected and executed.
There has been a lot of research done on the scheduling and routing of the long-distance vehicle. We propose a model for
long-distance vehicle transportation. Our model incorporates
the distance vehicle transport mechanism through agent-based
modeling. We checked our model to the basic design protocol of ABM
called Overview, Design, and Details (ODD). As we assume that
every category of agents is offering information as a service
we need a service-oriented protocol for the communication
between sensors and software components for monitoring and
tracking of vehicles on the route for better transportation. For
the service protocol, we incorporate concepts of the Internet of Things and Mobile ad-hoc networks \cite{raees2021context}.
Our proposed model incorporates the different services offered
by each entity. 

This section presented the introduction of our work. The remainder of this paper is organized as follows: In Section 2, we briefly describe some background studies for a better understanding of used concepts. Section 3 illustrates our ODD agent-based model. Section 4 describes a relevant scenario as we have made scenario-based validation. We conclude our paper in section 5.

\section{Background}
\label{section:2back}
Technology plays an important role in tracking and monitoring long-distance transportation. The term used for intelligent and connected devices is called Internet of Things. The Internet of Things (IoT) \cite{yang2010survey} has attracted the attention of both academia and industry. The European Commission has already predicted that by 2030, there will be more than 250 billion devices connected to the Internet \cite{sundmaeker2010vision}. IoT has the concept of being connected to people and things anytime, anyplace, with anything and anyone, ideally using any path/network and any service \cite{vermesan2022internet}.

The concept of offering Everything-as-a-Service (XaaS) \cite{banerjee2011everything} is a category of models introduced with cloud computing.
Vehicular systems with agent-based modeling have different applications of tracking and autonomy \cite{riaz2013lateral}, however, this case relates to tracking and management. 
Agent-based modeling (ABM) is used to model systems that are comprised of individual, autonomous, interacting “agents”.
ABM thus follows a bottom-up approach to understanding real-world systems \cite{macal2014introductory}.
A lot of work has been done on the tracking and monitoring of long-distance vehicles using real-time and passive systems. In
\cite{jog2016monitoring} a low-cost vehicle monitoring system is presented. Work states that the shipping industry has developed many tracking and monitoring systems first to determine where each vehicle was at any time. While in the modeling perspective, there are many models predicted for long-distance vehicle transport. Recently, also several agents-based freight transport analysis models have been suggested, e.g., INTERLOG \cite{liedtke2009principles} and TAPAS \cite{holmgren2012tapas}, which belong to the class of micro-level models, where
individual entities are represented and the relations between
entities are typically studied over time.
In \cite{reis2014analysis} a running intermodal transport service was used as a case study. The performance of the inter-modal transport service was compared against a potential road transport service for a set of mode choice variables, including price, transit time, reliability, and flexibility.

\section{LRV Monitoring and Support Modeling}
\label{section:3method}
Agent-based modeling has been itself grown into a vast field so
it is difficult to keep all of the model’s characteristics.
Many descriptions of Agent-Based Modelings presented in the literature are not complete, which makes it impossible to replicate
and re-implement the model. However, replication is key to science, and models should be reproduced. 
Also, the agent-based modeling descriptions are a lengthy mixture of words of factual description and these include long justifications, discussions, and explanations of many kinds. We have to read a lot about the model to understand even if the model itself is quite simple. 
The way to describe ABMs should be easy to understand yet it should describe a complete model. A known way to deal with such kind of problems is standardization. It is much easier to know and understand written material if the information is presented in a standardized way and we know the order of textual information. So a consistent protocol that is effective for Agent-Based Modeling becomes useful and it makes it easier to understand and write models.
A root-cause analysis can also provide an effective strategy to understand what to build in models \cite{raees2020study}.
To bring the benefits of standardization to ABMs, scientists have developed the ODD protocol for describing ABMs \cite{railsback2019agent}. “ODD” stands for “Overview, Design Concepts, and Details”: the protocol starts with three elements
that provide an overview of what the model is about and how it
is designed, followed by an element of design concepts that
depict the ABM’s essential characteristics, and it ends with
three elements that provide the details necessary to make the
description complete.

\begin{figure}[h]
  \centering
  \includegraphics[height=5cm]{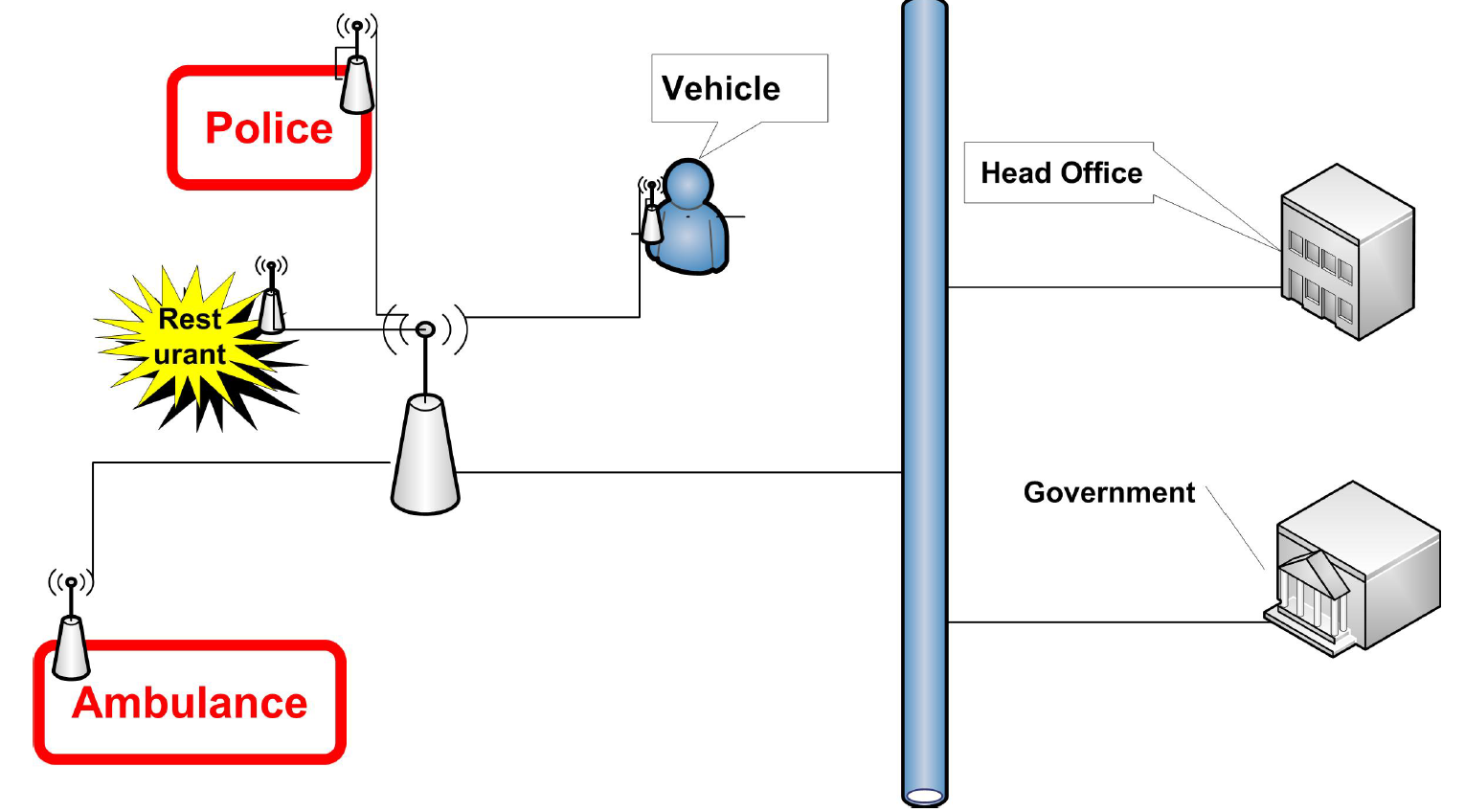}
  \caption{Conceptual Diagram of LRV Monitoring and Support Model.}
  \label{fig:fig1}
\end{figure}

\begin{figure}[h]
  \centering
  \includegraphics[height=6cm]{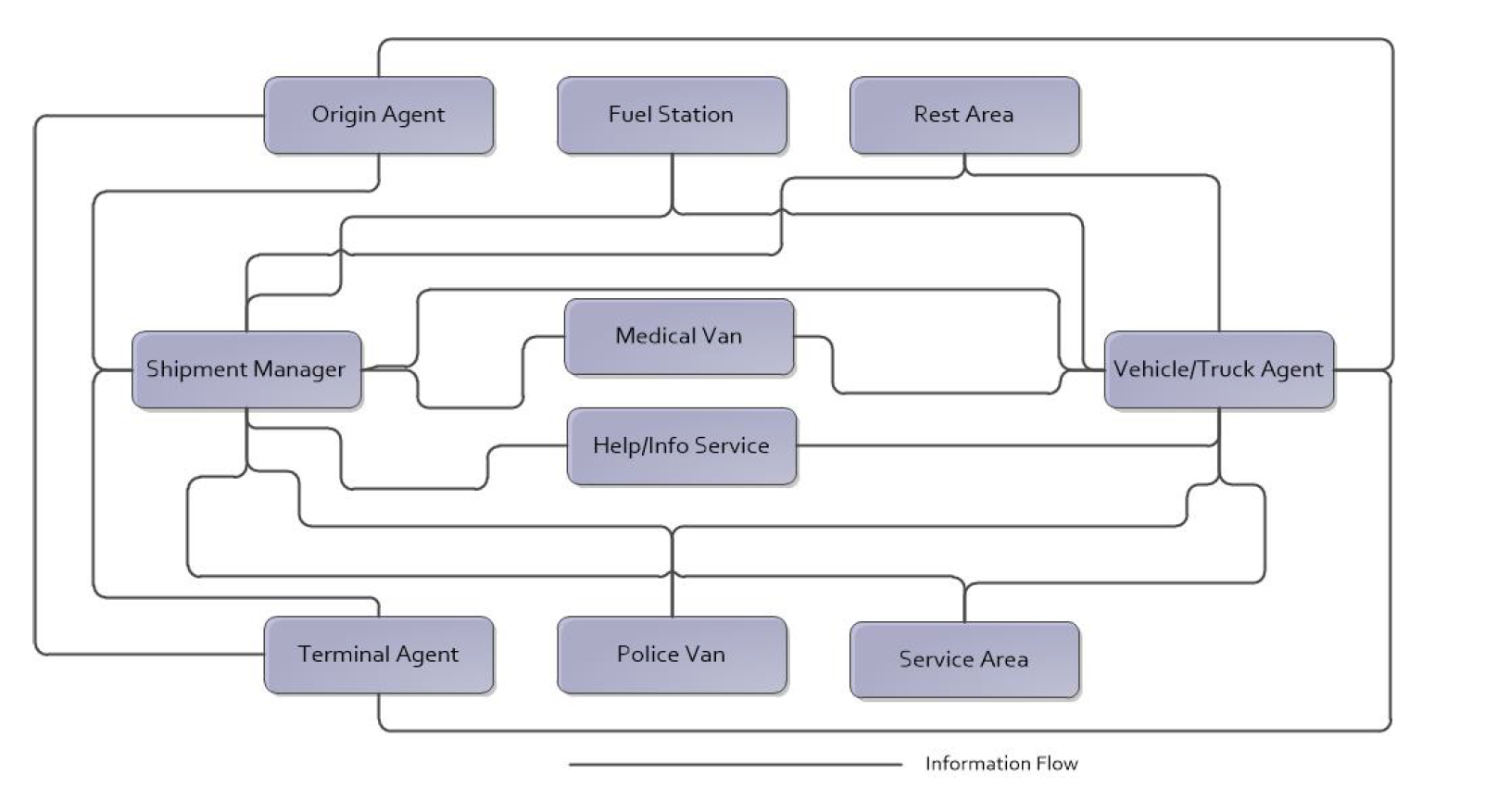}
  \caption{Conceptual Diagram of Agents Interaction.}
  \label{fig:fig2}
\end{figure}

\subsection{Agents}
After the analysis of the literature and problem domain, we identified agents that are involved in the interaction. Several agents were identified to model are shown in figure \ref{fig:fig1} and their interaction is shown in figure \ref{fig:fig2}.

\begin{itemize}
    \item Vehicle (trucks)
    \item Fuel station
    \item Rest Area
    \item Rescue/Medical Van
    \item Police Van
    \item Service Area
    \item Vehicle owner/manager company
    \item Origin agent (freight sender, shipment agent)
    \item Terminal agent (freight receiver)
\end{itemize}

\textbf{Origin Agents:} Origin agents are the agents that are the supplier/sender of goods. These are generated at the beginning of transport by booking with a transport company.

\textbf{Origin agents:} are the agents that are the
supplier/sender of goods. These are generated at the beginning of
transport by booking with a transport company.

\textbf{Owner company/Transport company:} Owner/transport
company is the central agent in the context as it is responsible for monitoring of vehicle. In case of any issue with the vehicle, the
transport company interacts with the police vans present near
the vehicle signal or in case of signal loss. These agents store
all the information about vehicle movements.

\textbf{Vehicle:} Vehicle agents are the actual trucks containing the goods. They interact with all other agents which help make a successful transport. The term we refer to connected as “connected smart objects”. Smart objects interact with each other throughout the journey and provide better decision-making for vehicles and other agents.

All other agents are helping agents in case of emergency or need. Interaction with a help service agent is assumed to be generalized for any kind of service at this point of modeling. As we make our model more concrete we will define the specific help service and its interaction with other agents.

\subsection{Behavioral modeling using ODD model}
\subsubsection{Overview}
\textbf{Purpose:} The purpose of the model is to design the movement of long-distance vehicle transportation, their interactions with the other helping agents, and find out the behaviors of agents over time. Under what circumstances does the vehicle need to contact police services, rescue services, rest areas, fuel stations, or any other agents? By understanding the need for any kind of service needed by a vehicle we can predict the time to reach destinations during any trip. We can efficiently provide information to vehicles in need of any kind of emergencies as well as the road and weather conditions to avoid any breakdowns. By gathering the data after applying the model we can predict why a vehicle takes more time to cover a distance. How often a vehicle needs interaction with other agents, and how these interactions will take place?

\textbf{Entities, state variables and scales:} The model has several kinds of entities including vehicles (trucks), shipper and receiver agents, help and control services, fuel stations, and rest areas. Each entity has some basic attributes to model for observing their behavior. To model the entities, we need to model roads and agents. Geographical road forms the patch areas. This can be read on a square grid for testing purposes.
For a real system, this is obtained as the location of any agent on
the ground. Each location has two variables longitude and latitude. Agents are described by which position they are in.
The vehicle has variables like speed, fuel capacity, reliability, travel time, load carrying, travel direction, vehicle category, etc. Police vans have variable coverage in which those can operate, it may be free or already engaged with some other agent. Same as for the rescue services. Fuel stations may indicate the level of fuel they have and their locations. Rest areas must be divided category-wise as well as the services they offer.

\textbf{Process overview and scheduling:} There are multiple processes in the model, the basic process is the movement of the vehicle (trucks) from one point to another. Interaction with other agents like police, service areas, ambulances, or any other service also forms a new process. However, interaction with other vehicles is somewhat important because there is no interaction with other vehicles. So, several processes are formed during vehicle transportation while interacting with other agents. Scheduling is also an important aspect of interaction processes in case of unavailability of an agent service.
There is a need to schedule the vehicle requests for services if an agent instance is not available temporarily or permanently.

\subsubsection{Design concept}
\textbf{Emergence:} The concept of emergence has been given connotations of being unexplainable in principle, but with ABMs, we are focusing on just the opposite: can our model system look and behave like the real one? The concepts that emerge from the model are not simply the sum of individual characteristics. Whether between any given two points same type of vehicle takes different times at the same load. The variability in the behavior of individuals makes some property to emerge from the model. The model's primary value is trip duration. Trip duration is counted as time taken from the starting point to the ending point.

\[
\text{TotalTime(actual)} = \sum_{i=0}^{n} \sum_{j=1}^{m} \Delta T_{i,j}
\]

Where total time is calculated as sum of time during each checkpoint up to n-number of checkpoint. During the trip number of checkpoints is defined for continuous monitoring of vehicle time. Typical actual/observed time duration between two checkpoints is defined as:

\[
\Delta T(\text{actual}) = 
\begin{cases} 
T_{\text{cp}}[j] - T_{\text{cp}}[i] & \text{if } i < j \\
\text{undefined} & \text{otherwise}
\end{cases}
\]

Tcp[j] is the time observed at a current checkpoint, and Tcp[i] is the time observed at a previous checkpoint. This observation is compared to average the time between two checkpoints and the number of parameters are updated depending on the total time consumed T and Dt by vehicle. If a vehicle is behind the schedule, route minimization is updated for vehicles such as taking short rest periods and increasing average speed within acceptable limits.
Important secondary values are reliability improvement, threats
faced and breakdown.

\textbf{Observation:} ABM can produce many kinds of dynamics \cite{raees2021context}. What we can learn depends upon what we observe from it. We need to observe each entity; and what individuals are doing at each time. So in our model, we observe each entity to know what they are doing at any time. These observations form a result that predicts the outcome of a specific process.

\textbf{Adaptive behavior:} The agents in this model adapt themselves during the trip. For example, if the weather forecast is bad for a specific area, the agent may rest early and take a longer trip later when the situation is better for traveling. Adaptive behavior is most needed in agent-based modeling and model entities must adapt to the best available solution at any given time.

\textbf{Sensing:} Sensing is an integral part of the agent-based models \cite{riaz2013lateral}. Vehicle agents with the help of the onboard unit should sense their current location and time to reach their next checkpoint. It is assumed that on onboard unit will calculate the time to reach till next checkpoint.
Certain help services offer their services to vehicles during the
trip to get better observations. By accessing the location of the vehicle, the location of the next checkpoint, and the average speed for the current area the time to reach the next checkpoint is calculated.

\[
\text{Time-to-reach-checkpoint} = \frac{Dv[j]}{AS[i,j]}
\]

Where Dv j is the distance between the vehicle’s current position v
and checkpoint j. ASi j is the average speed of the vehicle between checkpoint i and checkpoint j.

\textbf{Prediction:} Prediction is fundamental to decision-making. Prediction plays an important role in the evolution of our model. This cannot be done by the vehicle itself, rather it can be analyzed over time by the management company to help in better prediction of tours for a specific route. We can predict why a vehicle extra time in an area. We can predict the behavior of breakdowns while observing a vehicle and area’s breakdowns. With this kind of knowledge, we can make better decision-making and predictions. Let s be a sensor then the super choice will be

\[
S_s = \sup \{ s_1, s_2, s_3, \ldots, s_n \}
\]

Nearest Criteria choice will be

\[
N_s = \text{Nr}(s_1, s_2, s_3, \ldots, s_n)
\]

whereas, the best choice will be

\[
B_s = \text{Aggr} \left( B_s \bigcap N_s \right)
\]

\textbf{Interaction:} This model includes interaction with other agents like police vans, ambulances, and service areas in case of emergency. The vehicle must transmit a signal to the management company, police station, service area, or medical help in case of vehicle damage, some emergency, or threat. The vehicle must be contacted by the owner's company for further investigation. The other agents like police vans and ambulances must also be notified about the nature of the incident.

\textbf{Stochasticity:} Variability in the movement of the vehicle is too complex to represent. We may not be sure why a vehicle is taking longer time than expected in a certain area. The movement of agents cannot be the same for each trip within a specific area. This variability is represented by the reliability of the moving vehicle. The higher the reliability the higher the chances that the vehicle moves as per expectations through that area.

\subsubsection{Model Details}
\textbf{Initialization:} For initialization of the model we need to initialize the landscape upon which the vehicle has to travel. The initial population can be set in any range. We can increase or decrease the population. Initial parameters like vehicle reliability and service availability are set at the start.

\textbf{Input:} The model does not need any time series data.

\textbf{Sub models:} Models divided into further sub-models. A vehicle can use alternative paths during travel. The available services may not be available when needed also makes a sub-model that needs to be addressed.

\begin{figure}[h]
  \centering
  \includegraphics[height=8cm]{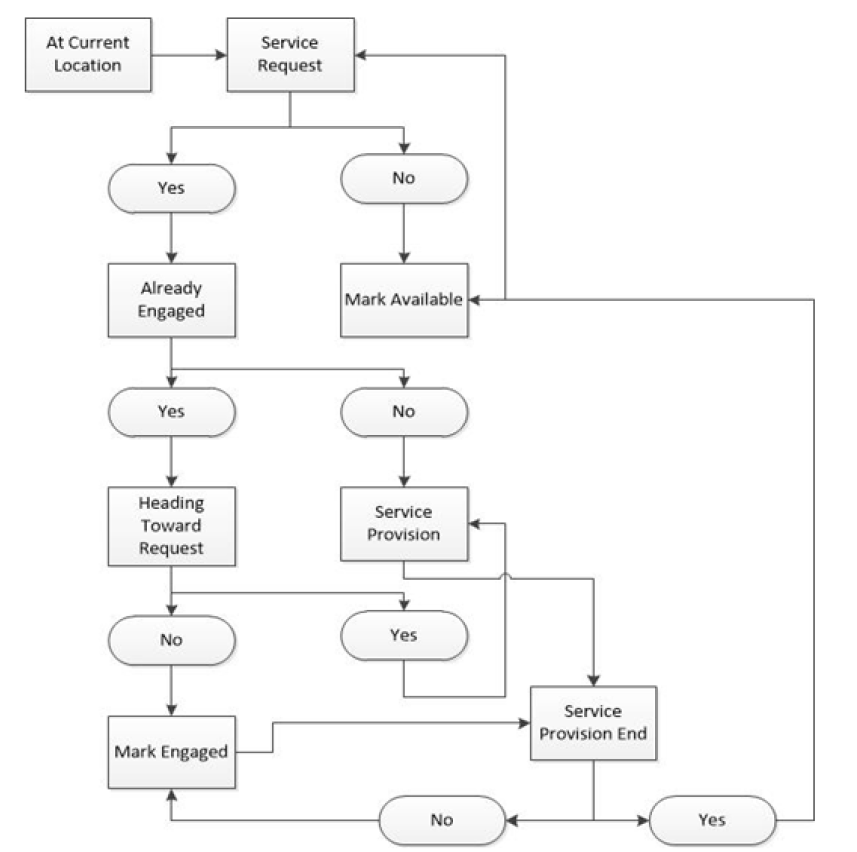}
  \caption{Police Van Service provision Sub-model.}
  \label{fig:fig3}
\end{figure}

\textbf{Police Service Model:} We break down the model into a set of sub-models, the interaction of a vehicle with the police vans forms a sub-model. Police vans are the agents that provide help services in case of danger of theft or accident during the trip. Each police van agent can communicate with vehicles within their predefined area and mark their service availability or unavailability. 
Figure \ref{fig:fig3} shows the simple service provision model for the interaction between a police van and vehicle agents. Police vans are the agents that are placed on the long-distance roads for the help of vehicles in case of emergency. Police vans are limited to cover a specific area assigned to them to offer services. Vehicles generate requests for the help service in an area. If the van does not have a pending request to process, it will mark itself as the available agent to offer its services. If the police van is already engaged with another assignment, it will check whether the new request is in the way to the current request the van is handling van can provide service in a way. If the new request is not in the current direction of the van it is to be marked engaged and the request is transferred to other vans. When the provision of service ends, the van will mark itself available for service.

\textbf{Fuel Station:} Vehicle agents can interact with the fuel stations on the road during the trip. Fuel stations can mark the availability of fuel to vehicles as well as prices and services available at stations.
Vehicles calculate the distance the reach the fuel stations and based on the current fuel level, the driver can assess whether to choose a station or travel to the next station for some more offered benefits as suits best.

\textbf{Service Areas/Medical Services:} Vehicle agents can interact with the service areas on the road during the trip in case of any fault in the vehicle. Agents may also seek medical help from the provided medical facilities on the road during the trip. Service area or medical services agents ask to offer help on-station or off-station depending upon the severity of the fault in a vehicle or need of medical assistance respectively.

\textbf{Shipment manager:} Interaction with the shipment manager is very important in our model. In case of any service needed or unavailability of service, the vehicle interacts with the shipment manager to seek help. Shipment managers can also interact with the help service agent depending on the location of the vehicle and the type of request.
Shipment managers can communicate with the help services so that they can better predict the actions needed to be taken by vehicle for on-time delivery. 

\textbf{Help Services:} Road conditions, weather situations, and traffic warnings play important roles in the successful completion of a trip. Sensing are reading these signals from provided sensors is very important for correct prediction of action to be taken.
Monitoring For monitoring purposes, we use the onboard units that must be installed on the vehicles to gather all the data about
vehicle movements.

\textbf{Use of the Internet of Things:} This model uses the recent trend of connected things to make an automated model of this process \cite{riaz2013lateral}. Internet of Things changing the way how things should be done. So we present a scenario of long-distance vehicle transport \cite{raees2021context}. In this scenario, we say a vehicle has to transport freight for a longer distance (say 2000 Km). The vehicle starts from the origin agent, by starting we mean the transport agent has information about the vehicle over a while so that he can predict the expected time to reach to destination. The expected arrival time is communicated to the vehicle and terminal agent. As the vehicle moves through the expected arrival time is updated upon every checkpoint/stop in the central server as travel history for future decisions. The weather and road situations are communicated to vehicles from time to time. In case of a signal lost from the vehicle, an alarm is generated at the transport agent's end and also on the police van present near the last signal received, so that the police vans are directed to check the related issue. In case of a medical emergency vehicle can generate an alarm to know about medical help available and ambulance services may be directed to the vehicle. Vehicles may also get knowledge of next rest points, service areas, and fuel stations to get a better view of their travel. Vehicle history is maintained during the whole journey; proper analysis of data can report the abnormalities to overcome bad encounters.
We may be able to know why a particular vehicle takes more time in an area while other vehicles don’t. why does one vehicle have more breakdowns than others? What is the vehicle's average time to reach
destinations? With this kind of knowledge, we can make better
decision-making and predictions.
Identifying such questions and digging deep to extract relevant answers is an essential technique \cite{raees2020study} which we may explore in the future in this context. 

\textbf{Protocols:} While taking everything as a service Interoperability is one of the major challenges in achieving the vision of the Internet of Things. The Semantic Gateway as Service provides a mechanism to integrate popular IoT application protocols to co-exist in a single gateway system. This protocol allows many of Internet of Things devices to connect with other web services. A study of different system web services is also needed.

\section{Scenario}
\label{section:4analysis}
Let’s consider the scenario of the China–Pakistan Economic Corridor of length 2;442 kilometers. After the completion of this corridor, China will use it for trading rather than South China
Sea route. If a loaded truck moves from Gawadar to Kashgar at an average speed of 50km/hour non-stop then, it will take 48:84 hours means 2:04 days. It’s impossible for a human to remain in a vehicle for such a long time. 
Normally, a truck driver requires 3 times meals a day and also some refreshments after every 3 to 4hours. Also, the truck will require refueling after a certain distance and we suppose it 5 times during this route.
Now, we calculate this time to reach the destination. We consider
every meal break of at least 30 minutes and refreshment break of 15 minutes and a fueling time of 15 minutes. Now, let’s calculate this time 

\[
\text{TotalTime} = 2:04 \times (3 \times 0.5) + 2:04 \times (6 \times 0.25) + 5 \times 0.25 + 48:84
\]

\[
\text{TotalTime} = 56:21 \text{ hours}
\]
\[
\text{TotalTime} = 2:34 \text{ days}
\]

So, after calculating this time by the model the company owning the fleet and the ultimate customer can expect when the truck will reach. Secondly, the truck driver can face any problem during this move. In case of any problem, he will inform the nearest police van directly, considering all the services are interconnected. In case the nearest police van is busy then in this scenario the system decides to convey the message to the second nearest van or wait for this to continue. So, this whole scenario is depicted in the model. 
Also, in case of an accident, the ambulance should be informed. And the most suitable ambulance should be assigned the duty.
In case of any mechanical problem, the mobile workshop should be informed. The same best workshop be selected based on the criteria. Also, the driver should be informed about the next filling station distance. Similarly, the driver should have the facility to check the next restaurant. The deals on the restaurant and rates of the restaurant should also be visible to the driver with the distance from the current position.
\section{Conclusion}
\label{section:6conclude}
Long-distance transport has both negative and positive effects on our economy and society. Monitoring and modeling long-distance transport has been found very interesting field by researchers. We have proposed a model here that uses the agent-based modeling approach to model this system as a multi-agent system. We have done some background study on how agent-based modeling works. Literature review shows that agent-based modeling approaches have been used to model transport systems lately. We propose our model and map with ABM protocol of Overview, Design concept, and Details (ODD). However, proper implementation of the model is still needed to be done. We propose the use of everything as a service concept, precisely sensing as a service where every agent provides services to others and uses the services of others. Service interoperability protocols have been studied. For services to connect between the Internet of Things and web services the Sematic Gateway as Service architecture has been found useful. The study of the most suitable protocol has been important work to focus on in the future.

 \bibliographystyle{elsarticle-num} 
 \bibliography{cas-refs}





\end{document}